\documentclass[12pt]{article}
\usepackage{graphicx}
\usepackage{epsfig}
\input babarsym

\begin{document}
\title{\bf Looking for Exotica at the \B\ Factories}
\author{Dr. Gagan Bihari Mohanty\\
University of Warwick, Coventry, UK\\
E-mail: mohanty@SLAC.Stanford.EDU\\
{\small Invited talk at the $17^{th}$ DAE-BRNS High Energy Physics Symposium,}\\
{\small IIT, Kharagpur, India, December 11--15, 2006.}}
\date{SLAC-PUB-12317\\
January 22, 2007}
\maketitle
\baselineskip=11pt

\begin{abstract}
Current experiments at the \B\ factories, designed to perform precision measurements of
matter-antimatter asymmetry in the \B\ meson system, have a much broader physics reach
especially in the sector of quarkonium spectroscopy. Here we present a minireview on
the new charmonium-like states observed at the \B\ factories including the $X(3872)$ and
$Y(4260)$.
\end{abstract}

\baselineskip=14pt

\section{Introduction}

More than 30 years after the November Revolution, quarkonium spectroscopy seems to
be on a great resurgence trail thanks to the discovery of many new states at the \B
factories and elsewhere. It all began with the discovery of the enigmatic $D_{sJ}(2317)$
resonance by the \babar\ Collaboration in April 2003\cite{babarDsj}. Soon after CLEO, Belle,
CDF and D\O\ at Fermilab, and BES joined the marathon raising the number almost to the
double digits. This collection of new states has in some cases confirmed quark model
and lattice QCD calculations while a few others poses serious challenge to our current
understanding of the strong interaction. In this report I present an experimental survey
of new charmonium-like states discovered and studied at the \B\ factories; the \babar\
experiment\cite{babar} at the PEP-II asymmetric energy $\epem$ storage rings located
at the Stanford Linear Accelerator Center (SLAC) in the US, and the Belle
experiment\cite{belle} at the KEKB energy-asymmetric $\epem$ collider in Tsukuba, Japan.

This minireview starts with some brief remarks on unique features of the experiments
those in conjunction with the high precision data delivered by the \B\ factories have
made this quarkonium sojourn a reality! The bulk of this report is concentrated on the
six new states, namely $X(3872)$, $X(3940)$, $Y(3940)$, $Y(4260)$, $Y(4320)$, and
$Z(3930)$ followed by possible theoretical model(s) to explain them. In the last
section I summarize my conclusions about these states.

\section{Experiments at the Beauty Factories}

Both \B-factory experiments, \babar\ and Belle are designed to carry out precision
$CP$ violation measurements on the beauty meson sector, so it is natural they have
so much in common. In particular, they have:
\begin{itemize}
\item A precision tracking system composed of a silicon vertex detector, whose main
      role is to locate the secondary decay vertices close to the origin, and a
      central drift chamber that measures momenta and angles of charged particles.
\item Excellent charged-particle identification (PID) capability accomplished by
      combining information on the specific ionization $(dE/dx)$ in the two
      tracking devices with that of a dedicated PID system. For the latter,
      \babar\ uses an innovative design in the Detector of Internally Reflected
      Cerenkov radiation. Belle makes use of an array of time-of-flight counters
      whose performance is augmented by an Aerogel Cerenkov Counter system at
      higher momenta.
\item An electromagnetic calorimeter, comprising CsI(Tl) crystals, which is used
      to measure the energies and angular positions of photons and electrons.
\item Iron flux return of the magnet instrumented with resistive plate chambers
      to identify muons and neutral hadrons over a broad kinematic range.
\end{itemize}
\begin{figure}
\begin{center}
\includegraphics[width=.48\columnwidth]{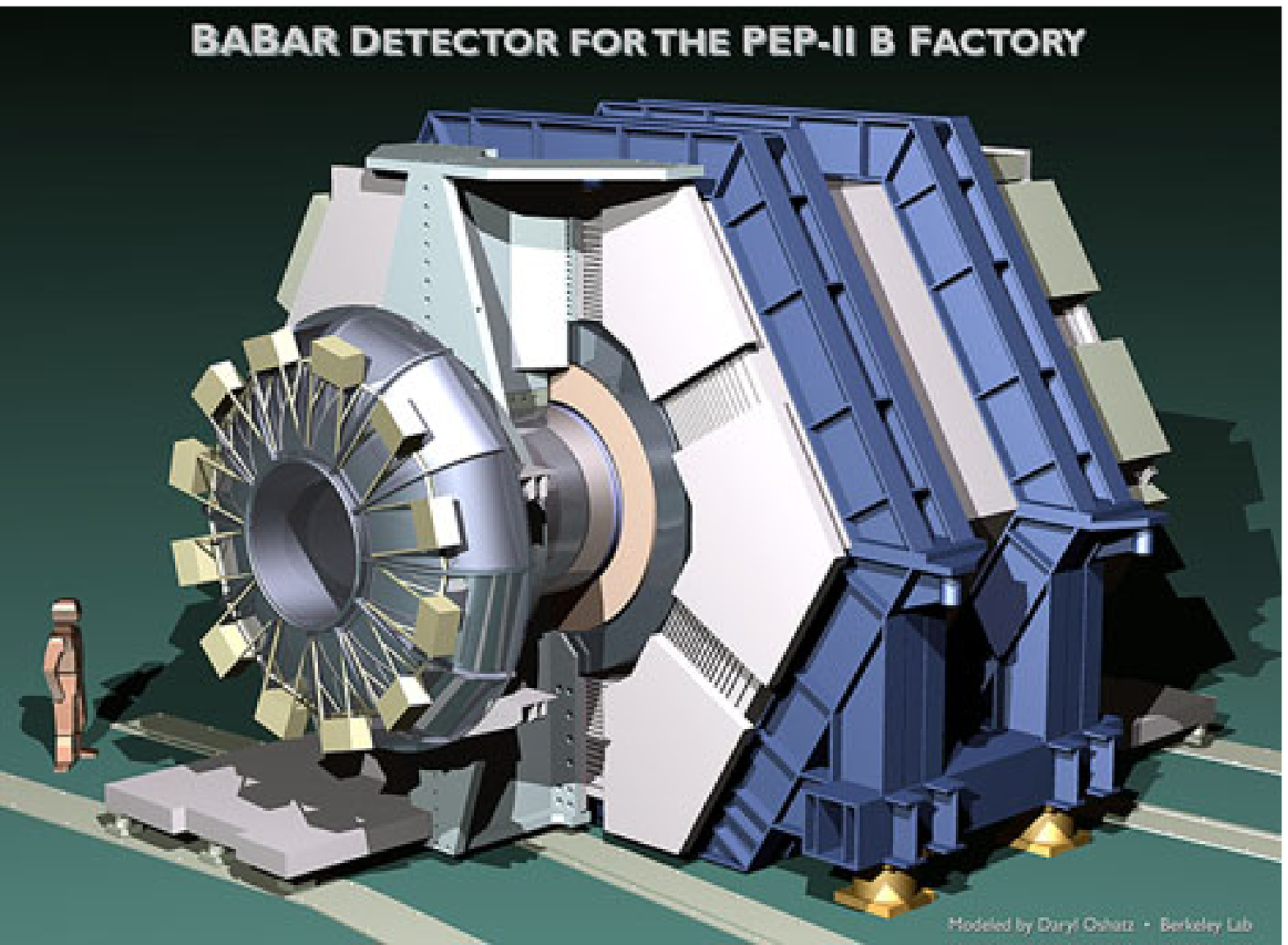}
\includegraphics[width=.51\columnwidth]{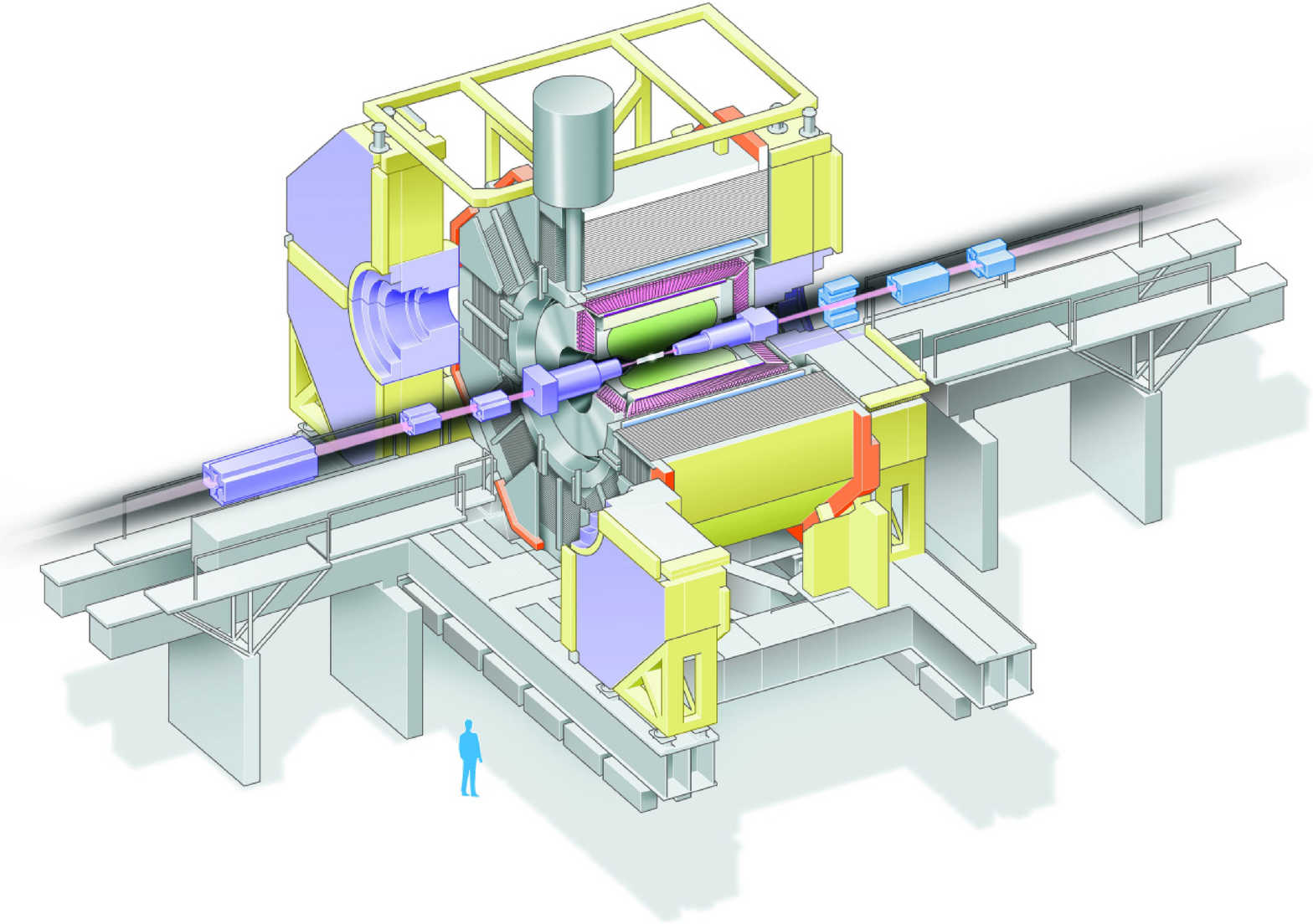}
\end{center}
\vspace*{-.3cm}
\caption{Layout of the \babar\ and Belle detectors.}
\label{fig0}
\end{figure} 
Figure~\ref{fig0} shows the layout of the \babar\ and Belle detectors. Although it
is the overall performance that drives the thing, the most important detector components
for spectroscopic studies are the PID system and the calorimetry. To add on to this
brilliant detector performance, the \B\ factories are accumulating high quality data
by leaps and bounds. The results reported here are based on data collected by \babar\ and
Belle although relevant results from other experiments will be cross-referenced as
appropriate.

\section{Ensemble of Exotica}

\subsection{{\bf{\it{X}}(3872)} - the poster boy}

The $X(3872)$ was first observed in the decay $B^-\ra X(3872)K^-,X\ra
J/\psi\,\pi^+\pi^-$\cite{charge} by the Belle Collaboration\cite{belleX1} and
subsequently confirmed by the \babar\cite{babarX1}, CDF\cite{cdfX1} and D\O\cite{d0X}
Collaborations. The mass of this state is $m=3871.2\pm 0.5$\,MeV and the width is
$\Gamma < 2.3$\,MeV at 90\,\% confidence level (CL) which is consistent with the
detector resolution. Study of the $\pi^+\pi^-$ invariant mass distribution in the
decay $X(3872)\ra J/\psi\,\pi^+\pi^-$ suggests that it may proceed through an intermediate
$\rho^0$ resonance. If so, one can expect to find its charged isospin partner $X(3872)^-$.
\babar\ has searched for this state in the decays $B^-\ra J/\psi\,\pi^-\pi^0K^0_S$
and $B^0\ra J/\psi\,\pi^-\pi^0K^+$. However, no evidence for a charged $X(3872)$ has
been found\cite{babarX2} and we set the following upper limits at 90\,\% CL:
$\mathcal{B}(B^-\ra X(3872)^-K^0_S,X^-\ra J/\psi\,\pi^-\pi^0)<11\times 10^{-6}$ and
$\mathcal{B}(B^0\ra X(3872)^-K^+,X^-\ra J/\psi\,\pi^-\pi^0)<5.4\times 10^{-6}$.
A search for $X(3872)$ in the initial state radiation (ISR) process $\epem\ra
X(3872)\g_{ISR}$ via the decay to $J/\psi\,\pi^+\pi^-$ has yielded a null
result\cite{babarX3}. This strongly disfavors a $J^{PC}=1^{--}$ assignment to $X$.
On the other hand, strong evidence of the radiative decay $X(3872)\ra J/\psi\,\g$
by both \babar\cite{babarX4} and Belle\cite{belleX3} experiments implies $C = +1$. This
when added together with angular analysis results from Belle\cite{belleX4} and
CDF\cite{cdfX2}, a $J^{PC}$ value of $1^{++}$ for the $X(3872)$ seems most likely. The
only known charmonium candidate that can come close to this pattern of measurements
is $\chi^\prime_{c1}$, but it has an expected mass of $3950$\,MeV. Clearly an
alternative approach beyond the traditional charmonium model is needed to explain
this state. The small width of the $X(3872)$ and its proximity to the $\Dz\Dstarzb$
threshold, $3871.8\pm 0.5$\,MeV, have led to the current two most-popular proposals;
\begin{figure}
\begin{center}
\includegraphics[width=.49\columnwidth]{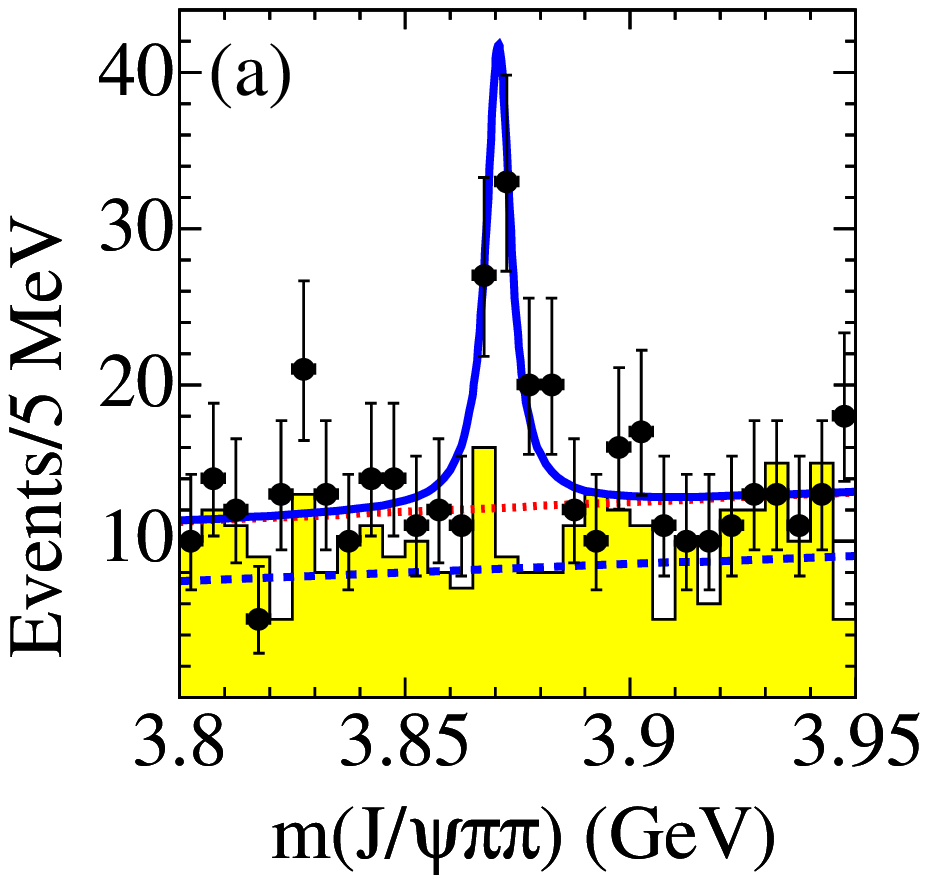}
\includegraphics[width=.49\columnwidth]{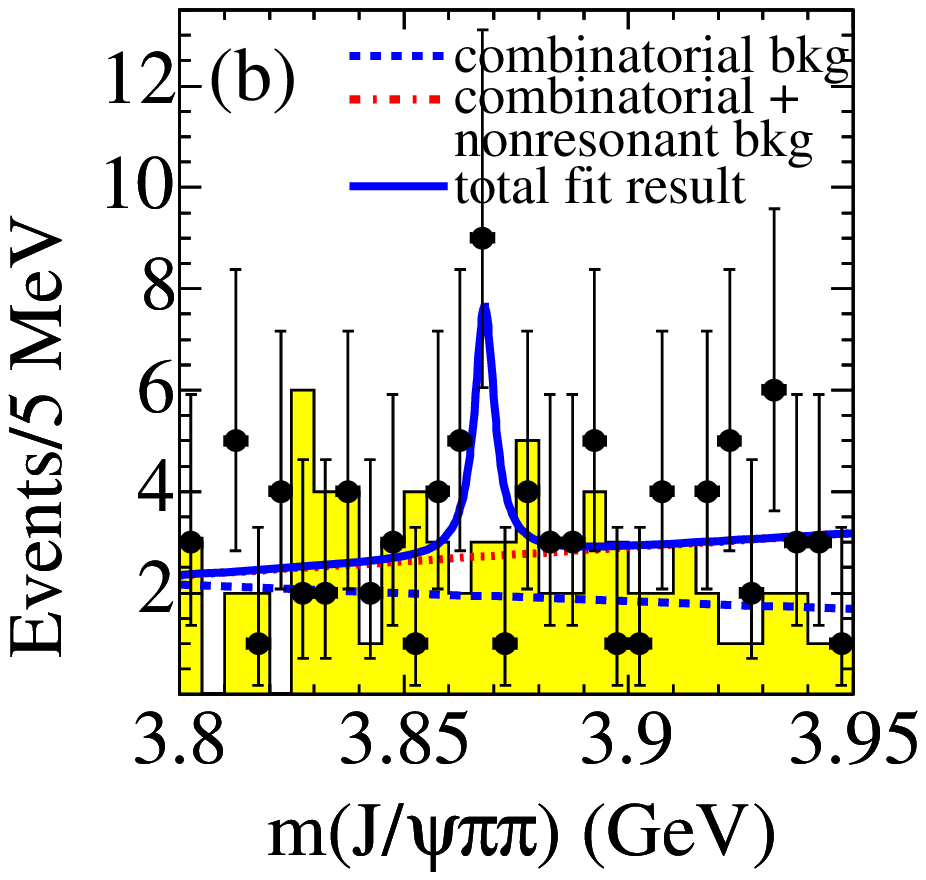}
\end{center}
\vspace*{-.3cm}
\caption{The $J/\psi\pi^+\pi^-$ invariant mass distribution in the signal region for
         (a) $B^- \ra X(3872)K^-$ and (b) $B^0 \ra X(3872)K^0_S$ decays from \babar.}
\label{fig1}
\end{figure} 
{\it i.e.} it could be a weakly bound $D\Dstarb$ molecule-like state\cite{theory1} or
a diquark-antidiquark state\cite{theory2}. To investigate these models, both \babar\ and
Belle have updated their earlier studies with higher statistics, and by looking into
other decay modes.

Using a dataset comprising about 232 million \BB\ pairs, \babar\ has studied
the decays $B^-\ra J/\psi\,\pi^+\pi^-K^-$ and $B^0\ra J/\psi\,\pi^+\pi^-K^0_S$.
Figure~\ref{fig1} shows the $J/\psi\,\pi^+\pi^-$ invariant mass distributions for
these two \B\ decay modes\cite{babarX5} which are fitted to a nonrelativistic
Breit-Wigner shape for the signal, and to a linear function for the nonresonant
and combinatorial backgrounds. For the charged \B\ mode, we obtain $61.2\pm 15.3$
while for the $B^0$ mode only $8.3\pm 4.5$ signal events. These yields are then
translated to the respective branching fractions by taking efficiency corrections
into account: ${\cal B}^-\equiv{\cal B}(B^-\ra X(3872)K^-,X\ra J/\psi\,\pi^+\pi^-)
=(10.1\pm 2.5\pm 1.0)\times 10^{-6}$ and ${\cal B}^0\equiv{\cal B}(B^0\ra X(3872)
K^0_S,X\ra J/\psi\,\pi^+\pi^-)=(5.1\pm 2.8\pm 0.7)\times 10^{-6}$, where
uncertainties are statistical and systematic, respectively. From these we derive
a ratio of the branching fractions, ${\cal R}={\cal B}^0/{\cal B}^-=0.50\pm 0.30\pm
0.05$. We also measure the mass difference of the $X(3872)$ state from the charged
and neutral \B\ decay modes, $\Delta m$, to be $2.7\pm 1.3\pm 0.2$\,MeV.
The diquark-antidiquark model predicts ${\cal R}=1$ and $\Delta m$ to be
$7\pm 2$\,MeV. The expected ratio of branching fractions is consistent with our
measurement, $0.13<{\cal R}<1.10$ at 90\,\% CL, and the observed $\Delta m$ is
consistent with zero, and with the model prediction within two standard deviations
($\sigma$). This result seems to slightly disfavor the molecule model that predicts
${\cal R}$ to be at most 10\,\%.

\begin{figure}
\begin{center}
\includegraphics[width=.98\columnwidth]{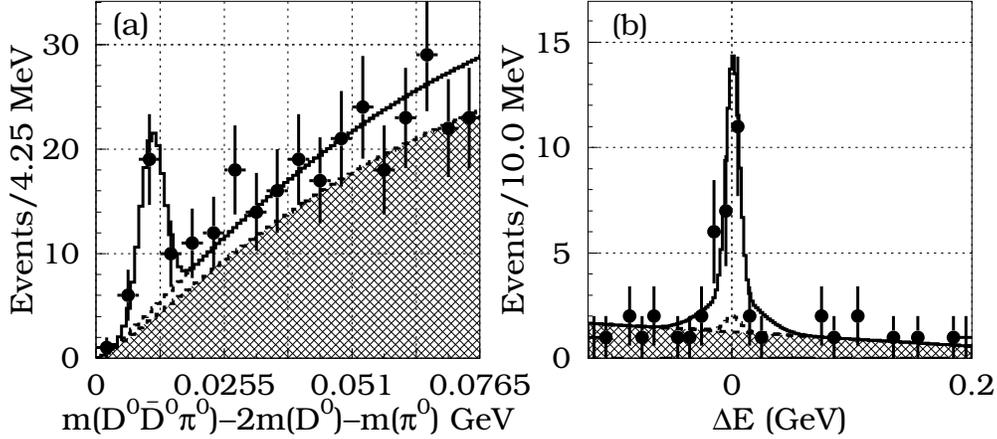}
\end{center}
\vspace*{-.3cm}
\caption{The projections of (a) $Q$-value and (b) $\Delta E$ distribution in the signal
         region from Belle. Dots are the data points, the hatched histogram corresponds
         to the combinatorial background; the dashed histogram denotes the total
         background while the solid line is the combined fit result.}
\label{fig1b}
\end{figure}

Belle has reported the observation of a near-threshold enhancement in the
$\Dz\Dzb\pi^0$ system from both neutral and charged decays of
$B\ra \Dz\Dzb\pi^0K$\cite{belleX5}. To determine the exact peak position as well
as the branching fraction, the two-dimensional distribution of $\Delta E$ {\it i.e.}
the difference of energy of the \B\ candidate and the beam energy {\it vs.} $Q$-value
($=m_{\Dz\Dzb\pi^0}-2m_{\Dz}-m_{\pi^0}$) is fitted. Projections onto $Q$-value
for $|\Delta E|<25$\,MeV and $\Delta E$ for $6$\,MeV $<Q$-value $<14$\,MeV are shown
in Figure~\ref{fig1b} along with the total fit result. The observed $\Dz\Dzb\pi^0$
mass is found out to be $3875.2\pm 0.7^{+0.3}_{-1.6}\pm 0.8$\,MeV,
where the first error is statistical, the asymmetric one is due to uncertainty on
the $\pi^0$ energy scale and the last one is due to the uncertainty associated with
the $\Dz$ mass. Assuming the threshold peak is due to the production and subsequent
decay of $X(3872)$ to $\Dz\Dzb\pi^0$ final state, this measurement is 2\,$\sigma$
higher than the world-average value of $X(3872)$ mass reported earlier. Corresponding
branching fractions for the neutral and charged modes are
${\cal B}(B^0\ra X(3872)K^0_{S},X\ra\Dz\Dzb\pi^0)=(1.66\pm 0.70^{+0.32}_{-0.37})
\times 10^{-4}$ and ${\cal B}(B^-\ra X(3872)K^-,X\ra\Dz\Dzb\pi^0)=(1.02\pm
0.31^{+0.21}_{-0.29})\times 10^{-4}$. This can be translated to ${\cal R}$ (ratio of
branching fractions) to be little over one which is in reasonable agreement with the
diquark-antidiquark model. However, we need more data to discriminate convincingly
between these two models.

\subsection{{\bf{\it{X}}(3940)} in double charmonium}

At the \B\ factories exclusive \B\ decays are not the only source of charmonium
states. \babar\ has also studied double charmonium production in the process
$\epem\ra\g^*\ra J/\psi\,{c\cbar}$ using 124\,${\rm fb}^{-1}$ of data\cite{babarCC}.
Only ${c\cbar}$ states with even C-parity are expected in this reaction, although
if there is a contribution from $\epem\ra\g^*\g^*\ra J/\psi\,{c\cbar}$, odd C-parity
states could also be produced. In this study, we first reconstruct a $J/\psi$
candidate via the $\epem$ or $\mumu$ decay mode and then calculate the mass of
the system recoiling against it. Three even C-parity charmonium states, $\eta_c$,
$\chi_{c0}$, and $\eta_c(2S)$ are observed, while there is no evidence for any
odd C-parity state such as the $J/\psi$. The distribution is fitted to obtain
the yield for each state, from which the production cross section is calculated.
Due to the requirement of at least five charged tracks in the event for
background suppression purposes, we report the product of the production cross
section and the branching fraction to states with more than two tracks. The
results are $17.6\pm 2.8^{+1.5}_{-2.1}$\,fb, $10.3\pm 2.5^{+1.4}_{-1.8}$\,fb
and $16.4\pm 3.7^{+2.4}_{-3.0}$\,fb for $\eta_c$, $\chi_{c0}$ and $\eta_c(2S)$,
respectively. The Belle measurement\cite{belleCC} in these three charmonium states
agrees with \babar\, and both are an order of magnitude higher than those predicted
by nonrelativistic QCD\cite{theory3}. However, recent works incorporating charm
quark dynamics\cite{theory4} seem to narrow down the discrepancy.

\begin{figure}
\begin{center}
\includegraphics[width=.7\columnwidth]{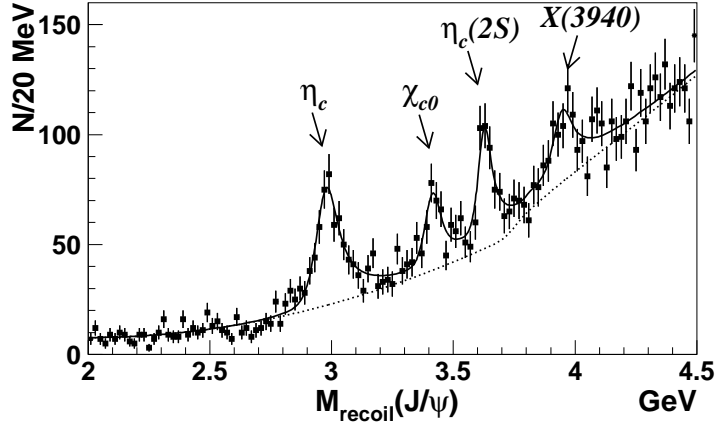}
\end{center}
\vspace*{-.3cm}
\caption{The $X(3940)$ signal in the $J/\psi$ recoil mass spectrum of the
         double charmonium production process from Belle. Black dots are data, the
         dashed curve is the background contribution while total fit result is denoted
         by the solid line.}
 \label{fig2}
\end{figure}

An interesting addendum to these measurements is the observation of a new
resonance by Belle\cite{belleCC} at a Breit-Wigner mass of $3943\pm 6\pm 6$\,MeV and
width $15.4\pm 10.1$\,MeV as shown in Figure~\ref{fig2}. The new state $X(3940)$ is
found to decay to $D\Dstarb$, but not to $D\Dbar$ or $\omega J/\psi$.
The dominance of the $D\Dstarb$ decay mode hints at the possibility that
the $X(3940)$ is the $2^3P_1({c\cbar})$ $\chi^\prime_{c1}$ state. It is natural
to try the $2P({c\cbar})$ since $2^3P_J$ states are predicted to lie in the
mass window of $[3920,3980]$\,MeV with expected widths $20-130$\,MeV. The
problem with this interpretation is, however, there is no clear-cut evidence
for the $2^3P_1({c\cbar})$ $\chi_{c1}$ in the plot. Thus one is bound to suspect
that the $\chi^\prime_{c1}$ should not be a strong signal in this reaction.
This has led to alternative speculations that $X(3940)$ could be the
radially-excited $\eta^{\prime\prime}_c$. Unfortunately this also has its
own problem as the expected $\eta^{\prime\prime}_c$ mass is $4040-4060$\,MeV,
approximately 100\,MeV higher than the measured value. An angular analysis of
the decay products will cast some light on the quantum numbers of $X(3940)$
and hence help on unambiguously identifying the state.

\subsection{The {\bf{\it{Y}}(3940)} - is it a twin?}

Belle claims the discovery of a second resonance at 3940\,MeV\cite{belleY3940}, this
time as a threshold enhancement in the $\omega J/\psi$ subsystem of the process
$B\ra K\omega J/\psi,\,\omega\ra\pi^+\pi^-\pi^0$. The reported mass and width are
$3943\pm 11({\rm stat})\pm 13({\rm syst})$\,MeV and $87\pm 22({\rm stat})\pm
26({\rm syst})$\,MeV, respectively. The new state is referred as $Y(3940)$ and
has not been seen in the decay mode $Y\ra D\Dbar$ or $D\Dstarb$. The mass and
width being similar to those of $X(3940)$ suggests $Y$ could be a radially
excited $P$-wave charmonium. However, the $\omega J/\psi$ decay mode is
peculiar in a sense that Belle measures ${\cal B}(B\ra KY(3940))
{\cal B}(Y\ra\omega J/\psi)=(7.1\pm 1.3\pm 3.1)\times 10^{-5}$ while we expect
${\cal B}(B\ra K\chi^\prime_{cJ})<{\cal B}(B\ra K\chi_{cJ})\{=(4\pm 1)\times 10^{-4}\}$.
So if we have to identify $Y(3940)$ with $\chi^\prime_{cJ}$, it would imply
${\cal B}(Y\ra\omega J/\psi)>12\,\%$, which is quite unusual for a canonical
${c\cbar}$ state above the open charm threshold.

Thus the $Y(3940)$ is something of an enigma, driving the claim of the Belle
collaboration that it is a charmonium hybrid. The large width coupled with the
unusual decay mode are consistent with the claim that a hybrid should have
strongly suppressed $D\Dbar$, $D\Dstarb$ or $\Dstar\Dstarb$ decay modes. However,
a mass less than 4000\,MeV is in conflict with the lattice QCD computations of
low-lying hybrid spectrum. Thus more data are required before any concrete
statement can be made on the exotic nature of $Y(3940)$.

\subsection{Observation of the $Y$(4260)}

\begin{figure}
\begin{center}
\includegraphics[width=.7\columnwidth]{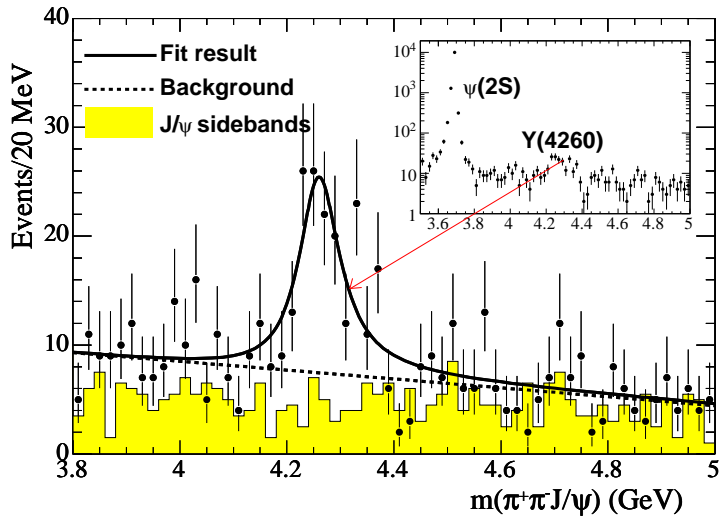}
\end{center}
\vspace*{-.3cm}
\caption{The $J/\psi\pi^+\pi^-$ invariant mass spectrum in the range 3.8$-$5.0\,GeV
         and (inset) over a wider range that includes the $\psi(2S)$ state from
         \babar.}
 \label{fig4}
\end{figure}

ISR events produced in the $\Upsilon(4S)$ energy region at the \B\ factories
serve to probe interesting physics occurring at lower center-of-mass energies.
Motivated by this, \babar\ has investigated the process $\epem\ra J/\psi\,\pi^+
\pi^-\g_{ISR}$ across the charmonium mass range, using a data sample of
233\,${\rm fb}^{-1}$ integrated luminosity\cite{babarY4260}. These events are
characterized by two pions, two leptons (electron or muon) making a $J/\psi$
candidate and a small recoil mass squared against the $J/\psi\,\pi^+\pi^-$ system.
Figure~\ref{fig4} shows the $J/\psi\,\pi^+\pi^-$ invariant mass spectrum for the
selected candidates. A clear enhancement near 4.2\,GeV is observed in addition
to the expected $\psi(2S)$ peak. No other structure is evident in the spectrum,
including the $X(3872)$. Using a maximum likelihood fit in which the signal is
described by a single relativistic Breit-Wigner lineshape, we obtain a yield of
$125\pm 23$ events with a statistical significance of 8\,$\sigma$ (the signal is
referred to as the $Y(4260)$). Assuming the peak is due to a single resonance,
the mass and width are determined to be $4259\pm 8^{+2}_{-6}$\,MeV and
$88\pm 23^{+6}_{-4}$\,MeV, respectively. We also calculate a value of
$\Gamma(Y(4260)\ra\epem)\cdot{\cal B}(Y\ra J/\psi\,\pi^+\pi^-)=5.5\pm
1.0^{+0.8}_{-0.7}$\,eV. Detection in the ISR process indicates $J^{PC} = 1^{--}$
for $Y(4260)$. We have weak evidence for $Y(4260)$ in the exclusive \B
decays\cite{babarX4}. In addition, it has been confirmed by the CLEO\cite{cleoY4260}
and Belle\cite{belleY4260} collaborations although at a slightly higher mass.

There are number of competing theoretical interpretations floating around the market to
explain $Y(4260)$, such as charmonium hybrid\cite{theory5}, hadronic molecule\cite{theory6}
{\it etc.}. However, we would need a multitude of measurements with higher
statistics to further pin down its properties, and hence the accompanying model.

\subsection{Hot of the press - the {\bf{\it{Y}}(4320)}?}

In an effort to clarify the nature of the $Y(4260)$, \babar\ has carried out a study of
the process $\epem\ra\psi(2S)\,\pi^+\pi^-,\psi(2S)\ra J/\psi\,\pi^+\pi^-$ using ISR events
over a wide kinematic range from the threshold up to 8\,GeV\cite{babarY4320}. The data
represent an integrated luminosity of 272\,${\rm fb}^{-1}$ recorded at $\sqrt{s}=10.58$\,GeV,
near the $\Upsilon(4S)$ resonance, and 26\,${\rm fb}^{-1}$ recorded near $10.54$\,GeV.
First a $J/\psi$ candidate is reconstructed through the decay $J/\psi\ra\ellell$
($\ell=e$ or $\mu)$ which is combined with a pair of pion tracks to form $\psi(2S)$.
Another pair of pions is then combined with the $\psi(2S)$ candidate and the resultant
invariant mass distribution is studied. The detection of the accompanying ISR photon is
not explicitly enforced as it is preferentially produced along the beampipe. Later,
however, by studying the missing momentum direction we confirm the ISR characteristics
of the event.
\begin{figure}
\begin{center}
\includegraphics[width=.7\columnwidth]{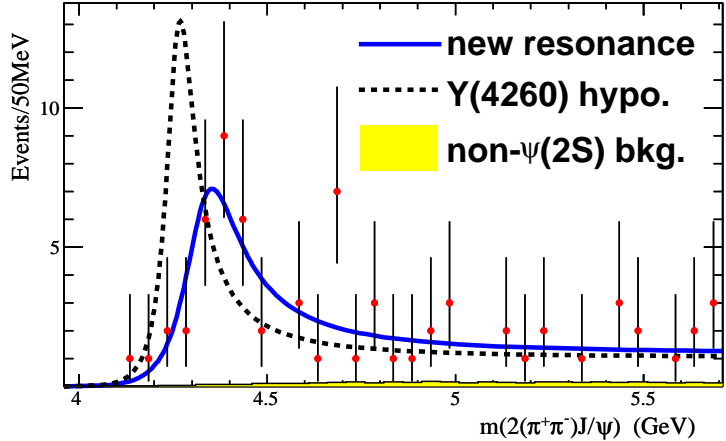}
\end{center}
\vspace*{-.3cm}
\caption{The $2(\pi^+\pi^-)J/\psi$ invariant mass spectrum up to 5.7\,GeV for the
         final sample from \babar. Red dots are data points while the shaded histogram
         and the curves are explained in the inset text.}
 \label{fig5}
\end{figure} 

Figure~\ref{fig5} shows the invariant mass distribution of the $2(\pi^+\pi^-)J/\psi$
system up to 5.7\,GeV in the final data sample. A structure around 4.32\,GeV is
observed in the mass spectrum. To better understand this structure, we fit the
full mass region under several hypotheses assuming the peak is due to: (a) $Y(4260)$,
(b) $\psi(4415)$ or (c) a new resonance whose mass and width kept as free parameters
in the fit. The probabilities of fit $\chi^2$ are determined to be $6.5\times 10^{-3}$,
$1.2\times 10^{-13}$ and 29\,\%, respectively. The new resonance hypothesis that
best describes the structure has a mass of $4324\pm 24$\,MeV and a width of
$172\pm 33$\,MeV, where the errors are statistical only. Given the close proximity
to $Y(4260)$ and statistical significance, more data would clarify the nature of
this structure.

\subsection{The {\bf{\it{Z}}(3930)} - the last alphabet?}

This state is observed by the Belle collaboration in the two-photon mediated process,
$\epem\ra\g^*\g^*\ra D\Dbar$\cite{belleZ3930} with a mass $3929\pm 5\pm 2$\,MeV and a
width $29\pm 10\pm 2$\,MeV where the quoted uncertainties are statistical and
systematic, respectively. The statistical significance of this claim is 5.5\,$\sigma$.
The $D\Dbar$ helicity distribution is consistent with the $J=2$ expectation. Similar
to $X(3940)$ and $Y(3940)$ the $Z$ seems an obvious candidate for $\chi^\prime_{c2}$.
The predicted mass and width of $\chi^\prime_{c2}$ are 3972\,MeV and 80\,MeV,
respectively. However, if we set the $\chi^\prime_{c2}$ mass to the measured value
of 3929\,MeV, its predicted width drops down to 47\,MeV (reasonably close to the
measurement) due to phase-space restriction. So at this stage we have no reason not
to believe that the $Z(3930)$ is the previously unseen $\chi^\prime_{c2}$ charmonium
state.

\section{Conclusions}

The last few years have been very exciting for hadron spectroscopy studies
at the \B\ factories. Both \babar\ and Belle are pioneering several sensitive
searches for new charmonium states, some of the most recent ones are summarized
in this report. As it seems $X$s, $Y$s, and $Z$s are seriously challenging
our current understanding of QCD in the nonperturbative domain. Least of the
all, $X(3872)$ and $Y(4260)$ have made it clear that the simple quark model
won't work for them and we need to look beyond the known horizon!

\hspace*{.1\textheight}
\begin{center}{\large\bf ACKNOWLEDGMENTS}\end{center}

It is a pleasure to thank the organizers of the DAE-BRNS High Energy Physics
Symposium for their kind invitation to present a talk on such an exciting
area of physics. Despite my earnest attempt I owe an apology to all whose
work could not be reported here for lack of space. This work is supported in
part by the Particle Physics and Astronomy Research Council of the UK, and
the US Department of Energy under contract number DE-AC02-76SF00515.


\begin{thebibliography}{99}
\bibitem{charge}  Charge conjugate reactions are included implicitly throughout.
\end{thebibliography}
\end{document}